# Corrected Navier-Stokes equations for compressible flows

Xu Jinglei[1,2], Ma Dong[1,2], Liu Pengxin[3], Bi Lin[3], Yuan Xianxu[*3] & Chen longfei[*1,2]

**For gas flows, the Navier-Stokes (NS) equations are established by mathematically expressing conservations of mass, momentum and energy. The advantage of the NS equations over the Euler equations is that the NS equations have taken into account the viscous stress caused by the thermal motion of molecules. The viscous stress arises from applying Isaac Newton's second law to fluid motion, together with the assumption that the stress is proportional to the gradient of velocity[1]. Thus, the assumption is the only empirical element in the NS equations, and this is actually the reason why the NS equations perform poorly under special circumstances. For example, the NS equations cannot describe rarefied gas flows and shock structure. This work proposed a correction to the NS equations with an argument that the viscous stress is proportional to the gradient of momentum when the flow is under compression, with zero additional empirical parameters. For the first time, the NS equations have been capable of accurately solving shock structure and rarefied gas flows. In addition, even for perfect gas, the accuracy of the prediction of heat flux rate is greatly improved. The corrected NS equations can readily be used to improve the accuracy in the computation of flows with density variations which is common in nature.**

As is well known, the Navier-Stokes (NS) equations are the governing equations for viscous fluid motions. The NS equations have been completely established by

[1]Department of Energy and Power Engineering, Beihang University, Beijing 100191, China. [2]Hangzhou Innovation Institute Yuhang, Beihang University, Hangzhou 311100, China.[3]State Key Laboratory of Aerodynamics, China Aerodynamics Research and Development Center, Mianyang 621000, China

Stokes[2] in 1845, since when the equations have been applied in a wide range of disciplines, including mechanical, civil, chemical and biomedical engineering, geophysics, oceanography, meteorology, astrophysics, and biology. The NS equations are initially developed for incompressible flows but are directly employed as the governing equation of compressible flows. The compressible NS equations reads,

$$\frac{\partial \rho}{\partial t} + \frac{\partial (\rho u_i)}{\partial x_i} = 0 \quad (1)$$

$$\frac{\partial (\rho u_i)}{\partial t} + \frac{\partial (\rho u_i u_j)}{\partial x_j} = \frac{\partial \sigma_{ij}}{\partial x_j} \quad (2)$$

$$\frac{\partial (\rho E)}{\partial t} + \frac{\partial (\rho E u_i)}{\partial x_j} = \frac{\partial (\sigma_{ij} u_j)}{\partial x_j} - \frac{\partial q_i}{\partial x_j} \quad (3)$$

where $\rho$, $u_i$, $\sigma_{ij}$, $E$ and $q_i$ are density, velocity, stress tensor, total energy per unit mass and heat flux respectively. The closure of above equations needs a state equation. For example, $p = \rho R T$ is the state equation for perfect gas, where $p$ is pressure, $R$ is specific gas constant and $T$ is temperature. The constitutive relation of Newtonian fluids is

$$\sigma_{ij} = -p\delta_{ij} + \tau_{ij}, \quad \tau_{ij} = 2\mu(S_{ij} - \frac{1}{3}S_{kk}\delta_{ij}) + \zeta S_{kk}\delta_{ij}, \quad S_{ij} = \frac{1}{2}(\frac{\partial u_i}{\partial x_j} + \frac{\partial u_j}{\partial x_i}) \quad (4)$$

where $\tau_{ij}$ is the viscous stress tensor, $S_{ij}$ is the strain rate tensor and its trace $S_{kk} = \nabla \cdot \mathbf{u}$ represents the compressibility. $\mu$ is the shear viscosity and $\zeta$ is the bulk viscosity which is assumed to be zero by Stokes.

In the early times, kinetic theory and acoustic techniques have demonstrated that the Stokes' hypothesis is not appropriate except in some special cases of monatomic gases[3,4]. RAJAGOPAL[5] has found that the Stokes' hypothesis is not valid for any fluid, and this includes monatomic gases. In recent studies, different formulas of bulk viscosity have been proposed in the computations of normal shock structure[6-8], compressible boundary layer[9-12], turbulence combustion[13,14], toroidal shock wave focusing[15] and so on, yielding generally improved predictions. The application of bulk viscosity is one way to improve the performance of NS equations, yet the modelling essence of bulk viscosity induces empirical coefficients which vary from case to case.

**Theory**

As we shall show in this work, the failure of the current compressible NS equations in the literatures lies in that the constitutive relation of $\tau_{ij}$ in equation (4), remains unchanged from incompressible to compressible condition. Note that, equation (2) are commonly in conservation form rather than convective form, indicating that researchers have realized the governing equations should be the transport equations for the momentum rather than the velocity. On the other hand, the viscous stress tensor, $\tau_{ij}$, is associated with the resistance offered by two adjacent layers of the fluid to their relative motion. According to Newton's friction law, the stress equals to the viscous coefficient multiplies the relative motion rate. Suppose there are two streams of Newton fluid shearing with each other, generating two shear layers, as shown in Figure 1a and Figure 1b. The mean momentum of the lower stream is $\rho u$. The difference is that, the mean momentum of the upper stream is $(\rho + \Delta\rho)(u + \Delta u)$ in Figure 1a while the mean momentum of the upper stream is $\rho(u + \Delta u)$ in Figure 1b. The temperature of the two streams are identical, so are the molecular viscosity, $\mu$. According to the classic NS equations, the viscous shear stress on the interface is $\tau = \mu \dfrac{\partial u}{\partial y} = \mu \dfrac{\Delta u}{\Delta y}$ in both Figure 1a and Figure 1b. However, the upper stream in Figure 1a has larger density than that in Figure 1b and the rest flow parameters are the same. If the flow is under compression, larger density is supposed to cause higher viscous shear stress, which is against that obtained from classic NS equations. This paradox might be caused by using velocity to calculate viscous stress, since velocity is a kinematic variable which does not contain physical significance. Thus, we shall introduce momentum to the constitution relation of viscous stress tensor in which, the relative motion rate should be the gradient of the momentum rather than the velocity,

$$\tau_{ij} = 2\frac{\mu}{\rho}(\tilde{S}_{ij} - \frac{1}{3}\tilde{S}_{kk}\delta_{ij}) + \zeta\tilde{S}_{kk}\delta_{ij}, \quad \tilde{S}_{ij} = \frac{1}{2}(\frac{\partial(\rho u_i)}{\partial x_j} + \frac{\partial(\rho u_j)}{\partial x_i}) \qquad (5)$$

This work still employs Stokes' hypothesis to evaluate solely the relative motion rate defined in equation (5). The gradient of momentum depends on the reference coordinate system, whereas the gradient of velocity does not. Thus, the choice of the reference coordinate system is crucial. As a matter of fact, the compression of fluids always happens due to the presents of solid walls, indicating that the reference coordinate should be fixed on the solid walls the fluids encounter. When the fluids leave the solid walls, the flow experience expansion. On one hand, the fluid streams depart from the walls, and thus it is not necessary to fix the reference coordinate system on the walls. On the other hand, no compression towards solid walls are imposed on the fluid streams. The fluid streams are not forced to interact with each other, so the viscous stress tensor should be described by equation (4). The final formula of viscous stress tensor is,

$$\tau_{ij} = \mu(S_{ij} - \frac{1}{3}S_{kk}\delta_{ij})(1 + sign(S_{kk})) + \frac{\mu}{\rho}(\tilde{S}_{ij} - \frac{1}{3}\tilde{S}_{kk}\delta_{ij})(1 - sign(S_{kk})) \quad (6)$$

where $S_{ij}$ is defined in equation (4), $\tilde{S}_{ij}$ is defined in equation (5), $sign(x) = -1$ when x is negative, and $sign(x) = 1$ when x is not negative. The two viscous forces presented in equation (6) reflects the ancient Chinese philosophy of ying and yang. The bulk viscosity, $\zeta$, is set to zero according to Stokes hypothesis. The NS equations which employ equation (6) as the constitutive relation of viscous stress is named as YNS (Yin-yang Navier-Stokes).

a                                        b

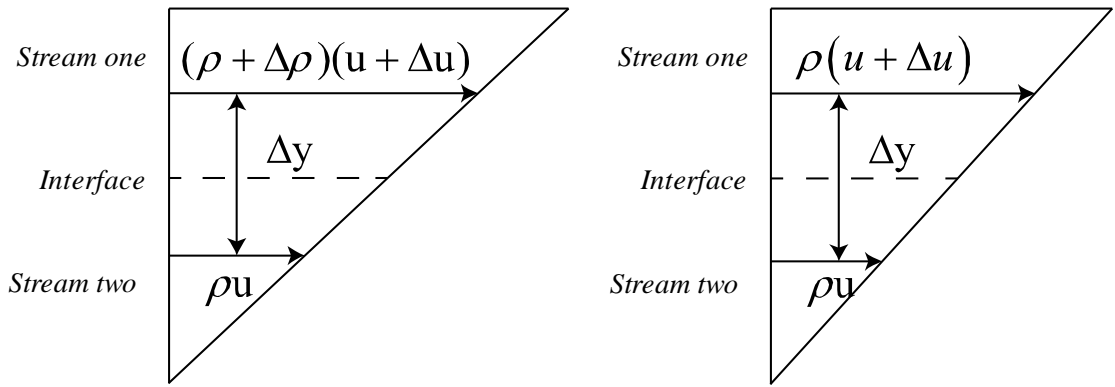

**Figure 1: viscous shear layer with (a) variable density and (b) constant density.**

**Applications**

**1 Shock structure**

The discrepancies between measured shock structure and predictions by the NS equations are commonly attribute to the Stokes' hypothesis. The shock structure[16] of $N_2$ is studied using the YNS equations. The normalized density profiles are shown in Figure 2a and Figure 2b. The density is normalized as $\rho = (\rho^* - \rho_1)/(\rho_2 - \rho_1)$ in which $\rho_1$ and $\rho_2$ are the densities in front of and behind the shock wave and $\rho^*$ is the local density. The length scale is normalized by the mean free path $\lambda$ in front of the shock wave. It is evident that compared with the NS equations, present equations better capture the density profile.

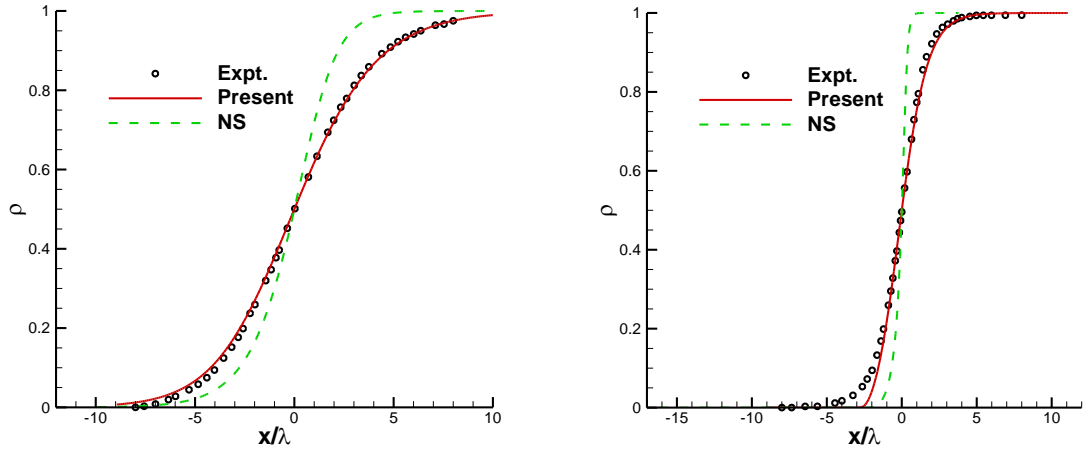

**Figure 2: normalized density profile in the shock wave for (a) Ma=1.53 and (b) Ma=10.00.**

## 2 Rarefied gas flow

The NS equations are conceived to be incapable of computing rarefied gas flows, and the community rely on the Boltzmann equations or DSMC method to compute such flows. In this section, hypersonic argon past a circular cylinder[17] is studied using both the NS and the YNS equations. The inflow temperature is 200K, the inflow density is $5.636 \times 10^{-6}$kg/m$^3$, the inflow Mach number is 10, the radius of the cylinder is 0.1524m, and the temperature on the surface of the cylinder is 500K. The Knudsen number which measures the degree of vacuum is Kn=0.05.

Figure 3 shows the computed heat flux coefficient on the surface of the cylinder. The heat flux coefficient, $C_h$, is defined as $C_h = Q/0.5\rho u^2$ in which $Q$ is the heat flux rate, $\rho$ is the inflow density and $u$ is the inflow velocity. $\theta$ is the angle (in degree) from the stagnation point. It is evident that the predictions of the YYNS equations agree well with that of the DSMC method, while the NS equations over-predict the heat flux.

The bow shock illustrated by the density gradient is shown in Figure 4. The NS equations yield two early a bow shock, and cause over-predicted heat flux on the surface of the cylinder.

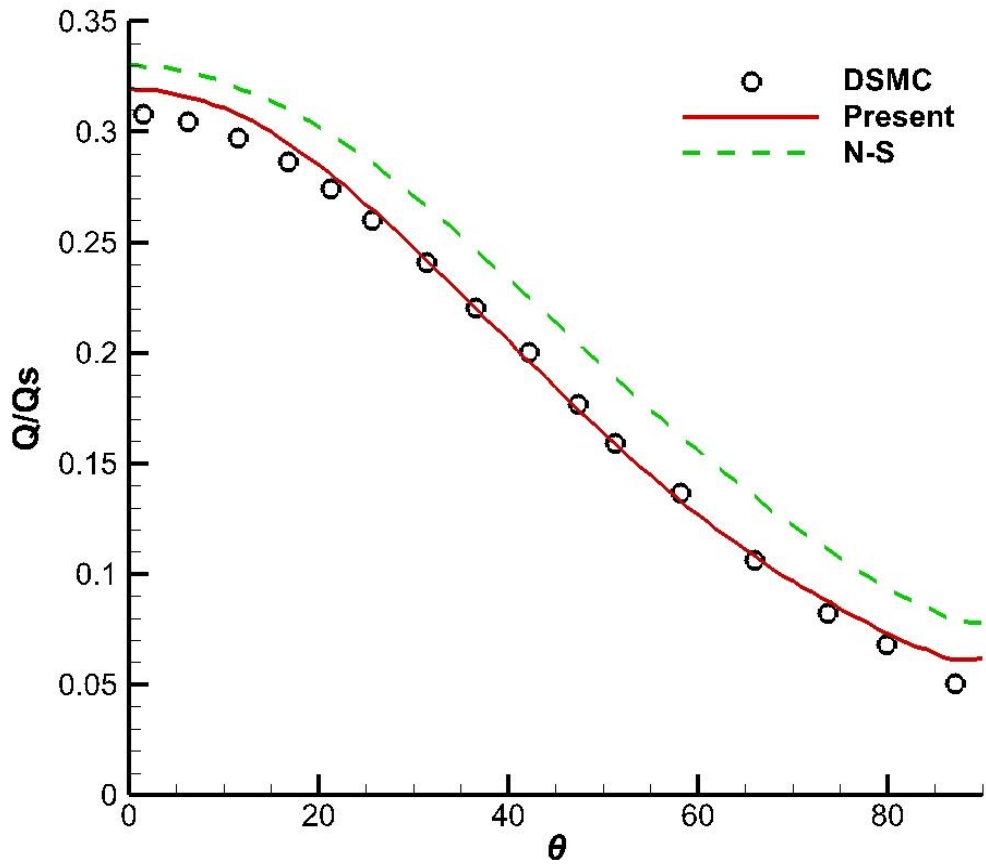

**Figure 3: Heat flux distribution on the surface of the cylinder.**

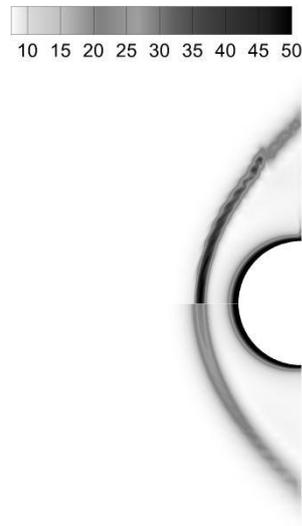

**Figure 4: Contours of $|\nabla\rho|/\rho$ predicted by the NS equations and the YYNS equations.**

### 3 Perfect gas flow

The blunt cone is employed as one typical geometry of the head of supersonic vehicles. The test case is taken from the NASA technical report[18], and the shock-wave predicted by YYNS equations is shown in Figure 5. The inflow conditions are $Ma = 10.6$, $T_\infty = 47.3K$, $Re_\infty/m = 3.937\times10^6$, $T_w = 294.44K$, and $\alpha = 0°$. The measured heat flux rate on the stagnation point is $215.8kw/m^2$, the computed value by the NS equations is $226.4kw/m^2$, and the computed value by the YYNS equations is $217.7kw/m^2$. The prediction errors are 4.9% for the NS equations and 0.9% for the YYNS equations, indicating that the new equations have made improvements on the predictions of stagnation heat flux rate.

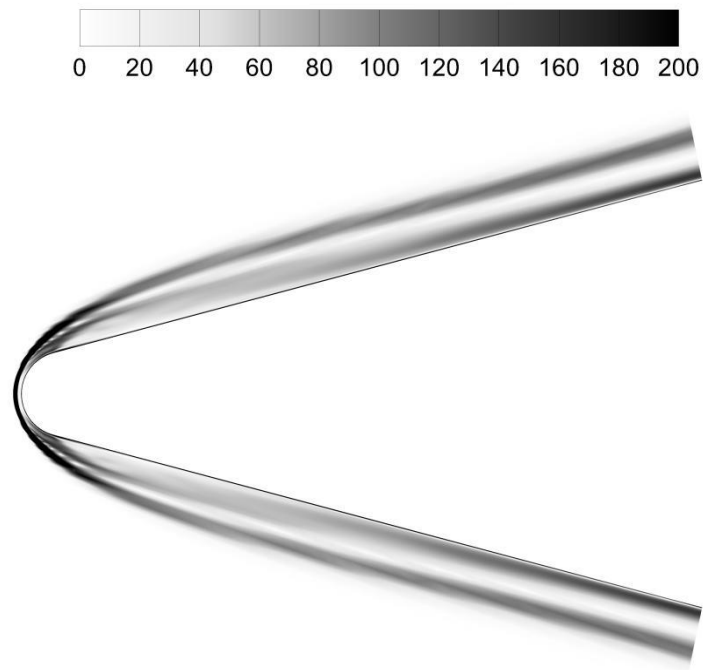

**Figure 5: Contours of |∇ρ|/ρ predicted by the YYNS equations.**

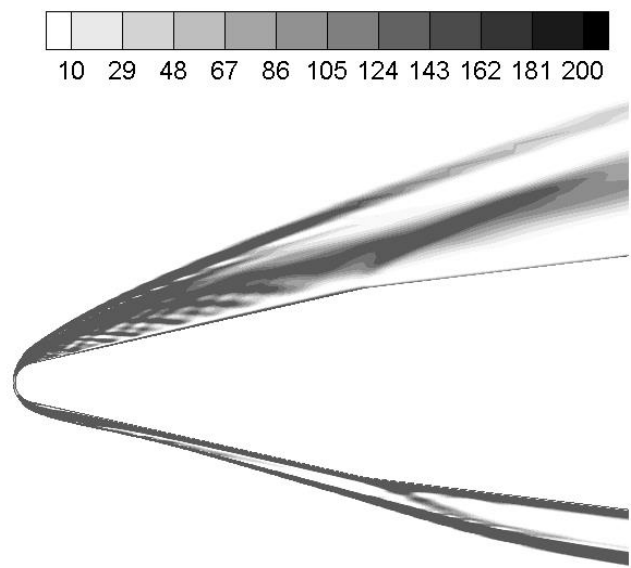

**Figure 6: Contours of |∇ρ|/ρ predicted by the YYNS equations.**

The above flows experience only compression. For the flow past a blunt double cone, the fluids at first experience compression and then experience expansion. The test case of blunt double cone is taken from another NASA technical report[19]. The flow conditions are Ma = 9.86, $P_\infty$ = 59.92 Pa, $T_\infty$ = 48.88K, $T_w$ = 300K, and α =10°. The heat flux coefficients on the windward side are shown in Figure 7. It is evident that for both the compression regions and the expansion regions, the YYNS equations again yield more accurate predictions.

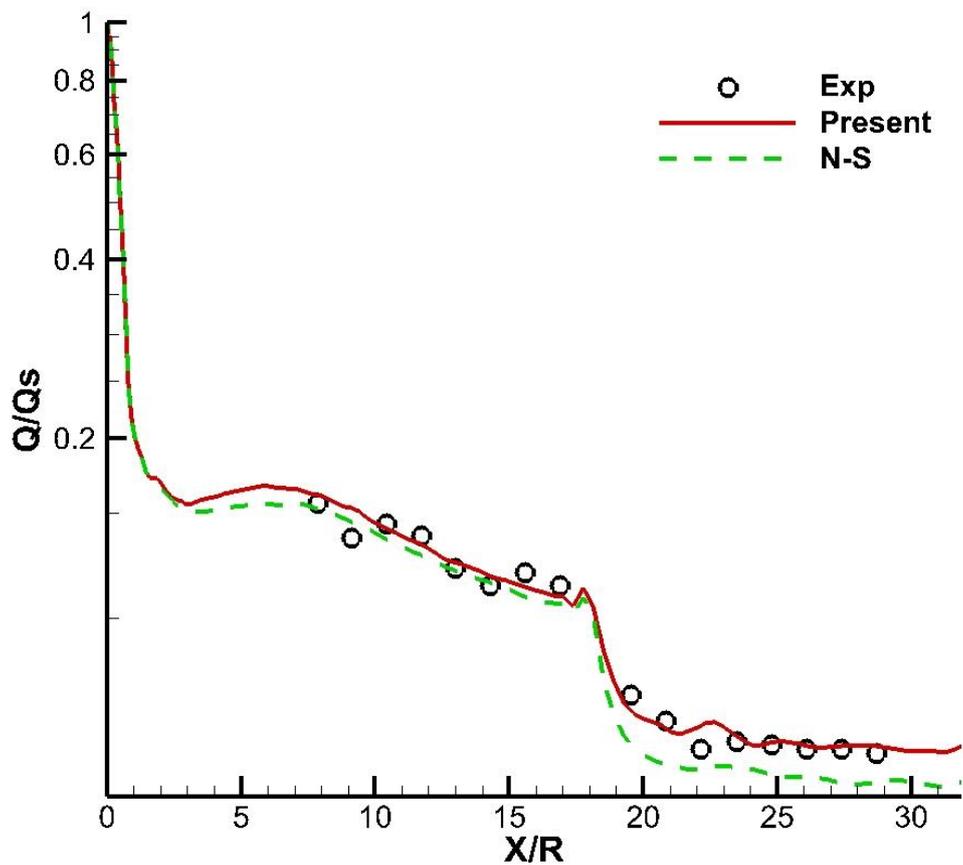

**Figure 7: Heat flux distribution on the windward side of the double cone.**

## References

1. https://en.wikipedia.org/wiki/Navier%E2%80%93Stokes_equations

**Acknowledgement** The authors would like to thank the National Numerical Wind tunnel Project (No. NNW2019ZT3-A14) and the Fundamental Research Funds for the Central Universities for supporting this research. The numerical calculations in this paper have received support from the super-computing system in the High-performance Computing Centre of Beihang Hangzhou Innovation Institute Yuhang.

**Author Information** Correspondence should be addressed to C.L. (chenlongfei@buaa.edu.cn) and Y.X.

(yuanxianxu@cardc.cn).